\documentclass{webofc}
\usepackage[varg]{txfonts}   
\begin{document}
%
\title{Many-probes multi-object spatially-resolved analyses of galaxy clusters in the big data era}
%
%

\author{\firstname{Fabio} \lastname{Castagna}\inst{1,2}\fnsep\thanks{\email{fabio.castagna@inaf.it}} \and 
	\firstname{Stefano} \lastname{Andreon}\inst{2} 
	\and
	\firstname{Alberto} \lastname{Trombetta}\inst{1}
	\and
	\firstname{Marco} \lastname{Landoni}\inst{3}
}

\institute{Department of Pure and Applied Sciences, Computer Science Division, Insubria University,
21100 Varese, Italy
\and
    INAF–Osservatorio Astronomico di Brera, via Brera 28, 20121 Milano, Italy
\and
    INAF-Osservatorio Astronomico di Brera, via Emilio Bianchi 46, 23807 Merate, Italy}

\newcommand{\jxs}{\texttt{JoXSZ\;}}
\newcommand{\jxsnosp}{\texttt{JoXSZ}}

\abstract{%
The thermal Sunyaev-Zeldovich (SZ) effect and the X-ray
emission offer separate and highly complementary probes of the thermodynamics of the intracluster medium, particularly on their radial dependence. We already released \jxsnosp, the first publicly available code designed to jointly fit SZ and X-ray data coming from various instruments to derive the thermodynamic radial profiles of galaxy clusters, including mass. \jxs follows a fully Bayesian forward-modelling approach, adopts flexible parametrization for the thermodynamic profiles and includes many useful options that users can customize according to their needs. We are including shear measurement in our joint analysis, and moving from single-cluster to multi-cluster analyses, allowing to quantify the heterogeneity of thermodynamic properties within the cluster population. At the same time, we are creating a suitable framework that efficiently stores and optimally processes huge volumes of data being released by the current and new generation surveys.
}
\maketitle
\section{Introduction}
\label{intro}

Galaxy clusters are the largest and most massive gravitationally bound objects in the Universe, and thus they offer a unique tracer of cosmic evolution \cite{Voit2005a}.
The thermodynamic properties of a galaxy cluster can be probed through observations in the optical, X-ray, or microwave bands via the Sunyaev-Zeldovich (SZ) effect \cite{Sunyaev1970, Sunyaev1972}, that we are combining into a single joint analysis.

We have already released \jxs \cite[][source code available on GitHub\footnote{https://github.com/fcastagna/joxsz}]{Castagna2020}, the first publicly available code designed to jointly fit SZ and X-ray data to derive the thermodynamic profiles of galaxy clusters.

We are currently working to include shear data as a third complementary component in the fit and to move from single-cluster to multi-cluster analyses, allowing to gather thermodynamic measurements on populations of clusters. Since it will involve large amount of observations coming from current and next generation facilities, we are creating a suitable framework that efficiently stores and optimally processes huge volumes of data.

\section{\jxs}
\label{sec-1}
\jxs is a Python code that combines SZ and X-ray data into a single analysis following a fully Bayesian forward-modelling approach. \jxs accounts for beam smearing and data analysis transfer function, for the SZ calibration uncertainty and X-ray and SZ background level systematics. It adopts extremely flexible parametrization for the thermodynamic profiles and it employs a consistent temperature across the various parts of the code, allowing for differences between X-ray and SZ gas mass weighted temperatures when required by the user, and calculates the correct Poisson-Gauss expression for the joint likelihood.

\jxs is built upon the SZ data fitting pipeline described in \texttt{PreProFit} \cite[][source code available on GitHub\footnote{https://github.com/fcastagna/preprofit}]{Castagna2019} and an updated version of the X-ray data cube fitter \texttt{MBProj2}, originally developed by \cite{Sanders2018}. \jxs merges and extends these two processes into a unique joint and consistent model based on a Markov chain Monte Carlo (MCMC) fitting algorithm.

\subsection{Inputs and Outputs}
For each cluster, \jxs requires the user to specify the SZ surface brightness radial profile, and the X-ray data cube in the form of multiple energy band radial profiles. 
The SZ analysis requires the point spread function (PSF), the SZ transfer function (a correction for data affected by filtering), and the conversion factor from Compton parameter to the observed data unit with temperature dependence (see \cite{Castagna2019} for more details). In X-ray, by using a data cube we exploit the information about temperature contained in the X-ray spectra. The components required for processing X-ray data, \textit{i.e.} the response matrix file (RMF) and the ancillary response file (ARF), must be provided by the user (see \cite{Castagna2020} for more details).

Through MCMC computation, \jxs samples the joint and marginal posterior distribution of parameters and derives the radial profiles of main thermodynamic properties, \textit{i.e}. pressure, electron density, temperature, entropy.
Other thermodynamic profiles are computed as usual \cite{Sanders2018}, including mass profile and gas fraction under the assumption of hydrostatic equilibrium.

\begin{figure}[h]
\begin{center}
\includegraphics[scale=0.5]{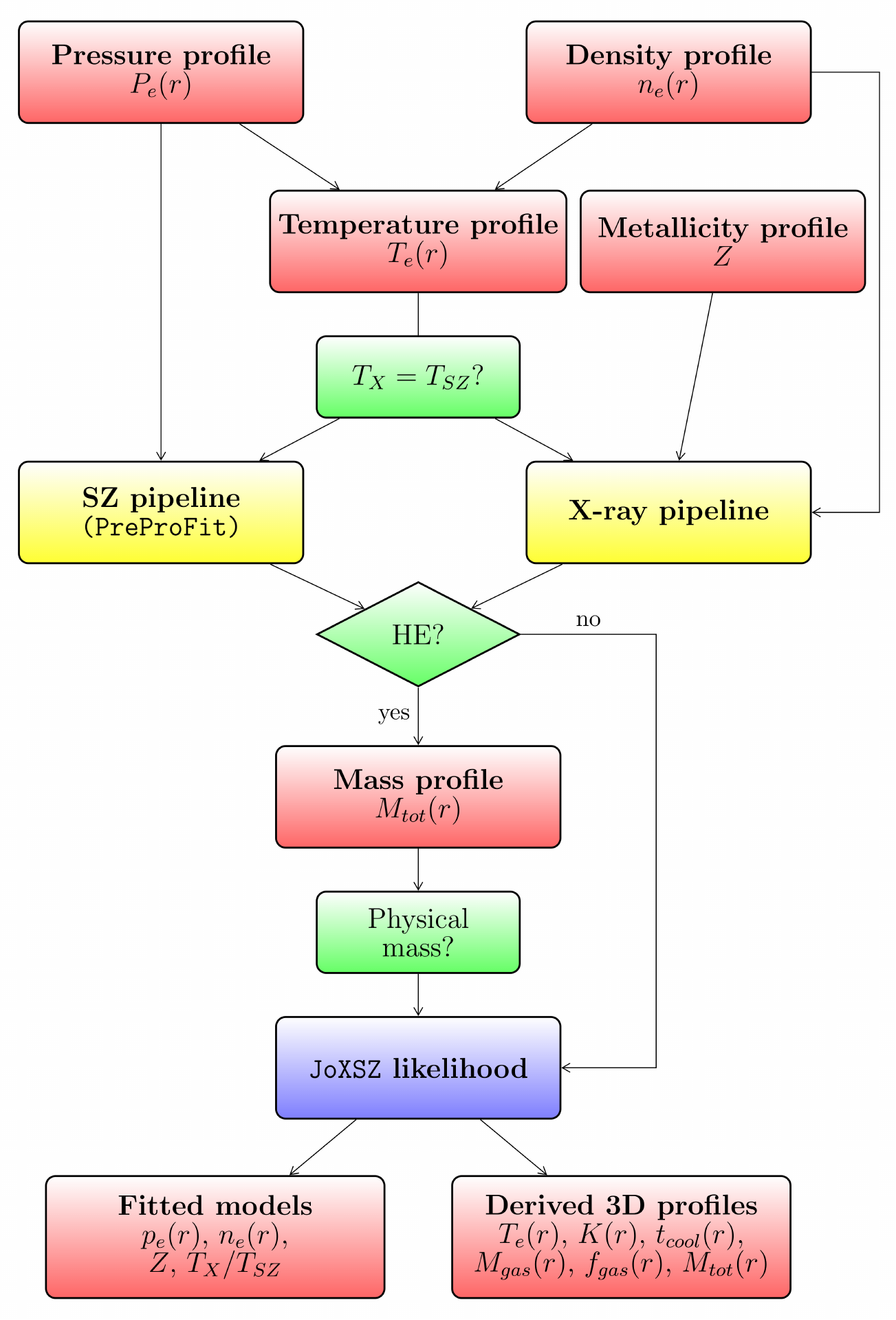}
\caption{Block diagram showing the program flow. Radial profiles are in red. Options are in green. Analysis pipelines are in yellow. Data enter in the blue box.}
\label{fig:fig_progflow}       
\end{center}
\end{figure}

\begin{figure}
\centering
\sidecaption
\includegraphics[width=9cm,clip]{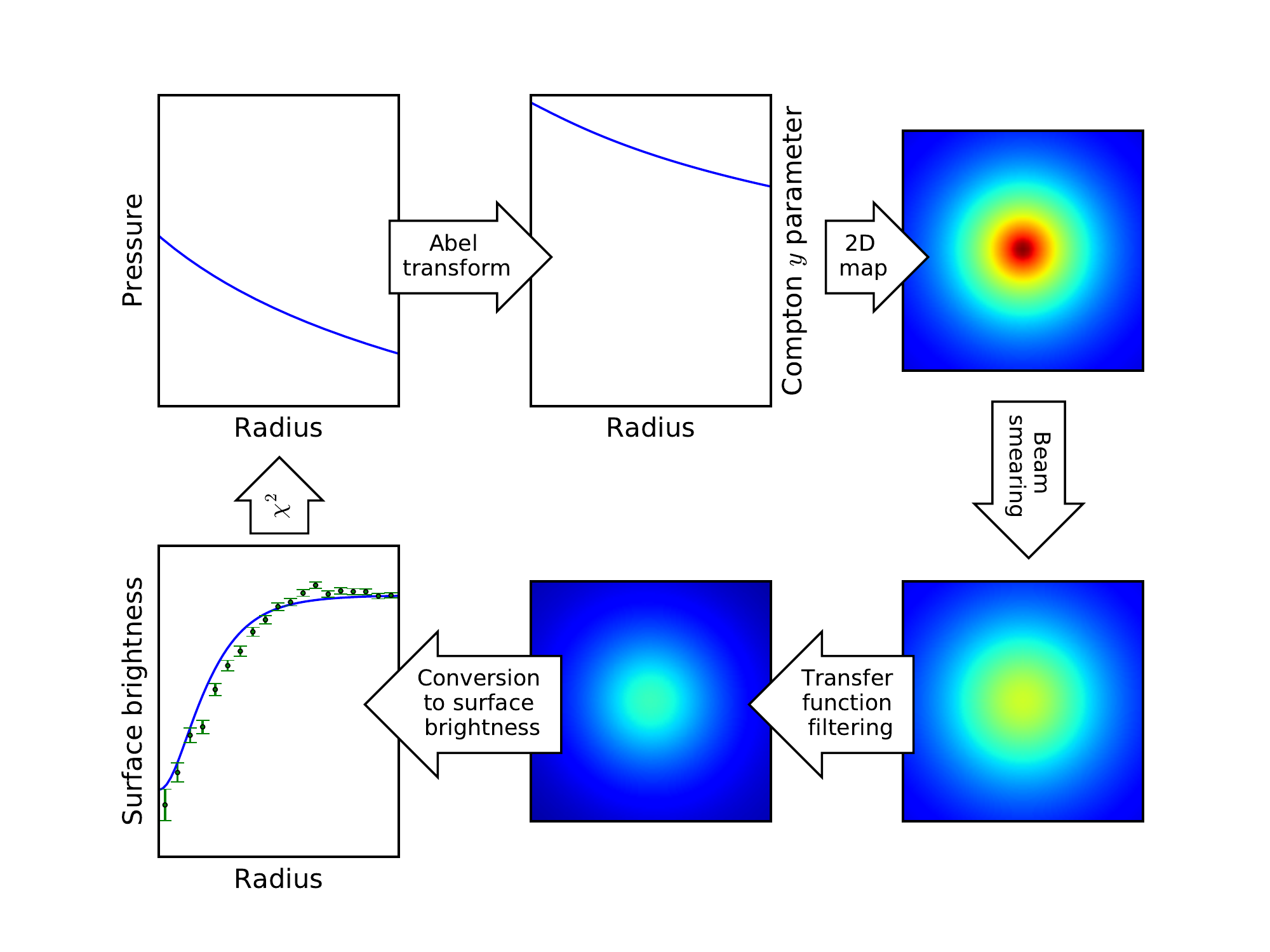}
\caption{Graphical representation of the SZ single-cluster analysis pipeline within \jxsnosp. The multi-object version loops over the clusters.}
\label{fig_sz}
\end{figure}

\begin{figure}
\centering
\sidecaption
\includegraphics[width=9cm,clip]{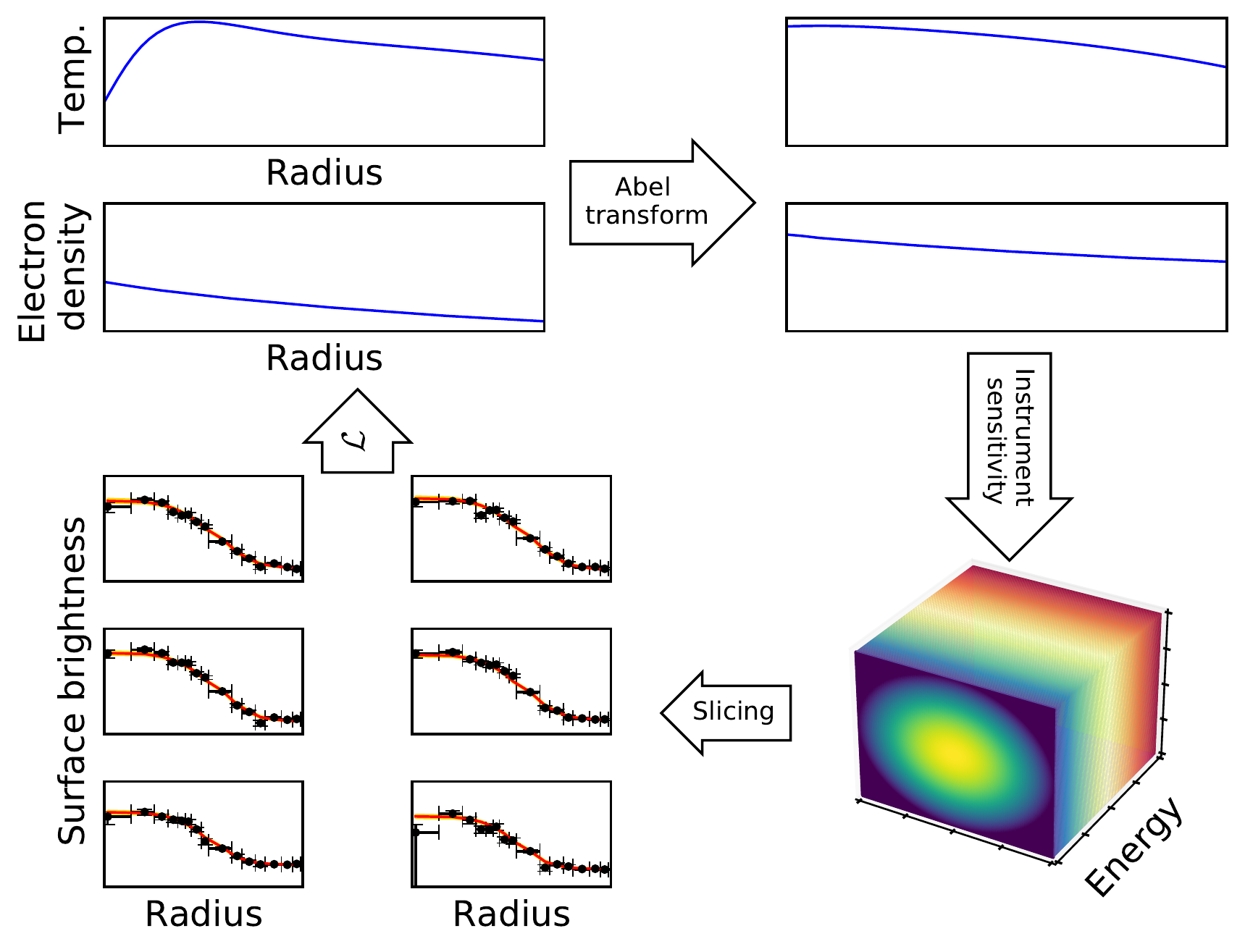}
\caption{Graphical representation of the X-ray single-cluster analysis pipeline within \jxsnosp. The multi-object version loops over the clusters.}
\label{fig_x}
\end{figure}

\subsection{Program flow}
Figure~\ref{fig:fig_progflow} depicts the block diagram describing the program flow of \jxsnosp. In the SZ pipeline, outlined in Fig.~\ref{fig_sz}, \jxs parametrizes the pressure profile, then projects it onto a two-dimensional map using forward Abel transform, convolves the map with the instrumental beam and the transfer function, and finally derives the surface brightness profile through appropriate conversion factors. If the beam file already includes the transfer function, the user has to specify it and \jxs will correctly handle it.
Regarding the X-ray analysis, shown in Fig.~\ref{fig_x}, \jxs makes full use of the spatial-spectral X-ray data cube, profiting of the better angular resolution of X-ray data, when available, to measure thermodynamic profiles at small radii unprobed by lower resolution SZ data.

\jxs is compatible with SZ data coming from different sources: NIKA, MUSTANG and SPT, whose provenence should be indicated because teams releasing the data adopts different standards. After specification, \jxs will process the data accordingly. \jxs has been already used for data coming from Chandra,  NIKA and MUSTANG \cite{Castagna2020, Andreon2021}, and is being used for data from XMM, XRT, SPT, and Planck \cite{Andreon2021prep}.
\jxs is also able to read data in different formats, and to automatically handle measurement units throughout the code using \texttt{astropy.units} \cite{Astropy2013} upon correct specification from the user of the measurement units of input data.

\subsection{Thermodynamic profiles modelization}
\label{sec-2}
Users are allowed to formulate the pressure profile through either parametric model or binned profile: a generalized Navarro, Frenk \& White (gNFW) model \cite{Nagai2007}, a cubic spline model or a power-law interpolation model. Optionally, users can apply a prior constraint on the outer slope of the pressure profile.
The multi-parameter (up to 10) model introduced by Vikhlinin \cite{Vikhlinin2006} is used
to parametrize the electronic density profile, and given the large number of possible
parameters it should be flexible enough to satisfy the needs of most users.

The temperature profile is simultaneously influenced by the pressure constraints from SZ observations and the 
X-ray data cube, \textit{i.e.} the X-ray spectra at various distances from the cluster center.
\jxs allows users either to consider a unique temperature profile $T_{SZ}=T_X$, or to make a distinction between the gas mass weighted temperature $T_{SZ}$ and the X-ray temperature $T_X$, introducing the multiplicative parameter $\log(T_X/T_{SZ})$.

\jxs assumes a flat metallicity profile, whose value can be fitted, or fixed,
at user request.

\subsection{Additional functionalities}
Additional functionalities are implemented in the code, each user can choose whether to adopt them or not. By default, \jxs includes a calibration parameter accounting for uncertainties in the SZ measurements and a X-ray background scaling parameter that accounts for differences in background level between the cluster and control field directions. In the SZ analysis, a pedestal parameter for the surface brightness can be added in the case of maps with non-zero level. Another option is to apply a prior constraint on the integrated Compton parameter.
\jxs can be easily used for feasibility studies adopting simulated data, e.g. Gaussian approximations for the PSF and the transfer function.

If dealing with a large filtering image, \jxs gives the opportunity to neglect extremely small scales to increase the execution
speed, but at the user's own risk.

To further save CPU time, users may ask to start the exploration of the parameter space not far from the likelihood maximum by specifying a (guessed) value for the cluster's characteristic radius,
from which pressure parameters are derived assuming 
the universal profile defined by Arnaud et al. \cite{Arnaud2010} as first
approximation. This option is also available for a binned pressure model: \jxs automatically converts the parameters from the universal profile into the required binning values.

\section{Current and future developments}

To fully take advantage of the multi-probe potential of current and incoming surveys, such as Vera Rubin Observatory, Euclid, eROSITA, Roman Space Telescope, ACT, and SPT, we are currently adding shear measurements to our joint SZ and X-ray analysis.
At the same time, we are paving the way to extend the analysis from single clusters at a time to multiple clusters at the same time, meaning that parameter estimation will be simultaneously conducted at both individual level and population of clusters level through a Bayesian hierarchical model.
Such a tool will be of great interest for the astrophysical community, giving the opportunity to answer more complex questions \cite{Feigelson2021}, e.g. testing General Relativity \cite{Cardone2021} or assessing the heterogeneity of thermodynamic properties within the cluster population, such as the scatter in pressure or mass profiles.

Since the code of \jxs includes highly time-consuming operations, we are currently working on optimizing the computation, both reducing redundant operations and exploiting parallel computing techniques. In view of the extension from single to multi-object analyses, we need to develop typical features of Big Data Analytics, 
in order to transform the huge avalanche of data into accurate information. To do so, we are constructing an adequate infrastructure which is based on state-of-the-art providers for data storage (Amazon Web Services, e.g. see Landoni et al. \cite{Landoni2019}) and data processing (Apache Spark \cite{zaharia2010}). 
Frameworks like Apache Spark allow us to define very specific functionalities on datasets of large dimensions, and to define computation procedures that can be easily scaled on parallel and distributed architectures. 
Furthermore, they are able to integrate data analysis techniques based on deep learning, such as convolution neural networks, recently used to estimate cluster masses through X-ray data.



%
\bibliography{references}
%
%
%
%

\end{document}